\documentclass[aps,prc,twocolumn,tightenlines,amssymb,aps,nobibnotes,superscriptaddress,showpacs,balancelastpage,floatfix]{revtex4-1}
\usepackage{graphicx}

\begin{document}


\title{
\begin{center}
Discovery of a new isomeric state in $^{68}$Ni:\\
Evidence for a highly-deformed proton intruder state.
\end{center}}


\author{A.~Dijon}
\affiliation {Grand Acc\'el\'erateur National d'Ions Lourds (GANIL), CEA/DSM-CNRS/IN2P3, Boulevard H. Becquerel, F-14076, Caen, France}
\author{E.~Cl\'ement}
\affiliation {Grand Acc\'el\'erateur National d'Ions Lourds (GANIL), CEA/DSM-CNRS/IN2P3, Boulevard H. Becquerel, F-14076, Caen, France}
\author{G.~de~France}
\affiliation {Grand Acc\'el\'erateur National d'Ions Lourds (GANIL), CEA/DSM-CNRS/IN2P3, Boulevard H. Becquerel, F-14076, Caen, France}
\author{G.~de~Angelis}
\affiliation {LNL(INFN), Laboratori Nazionali di Legnaro, I-35020 Legnaro, Italy}
\author{G.~Duch\^ene}
\affiliation {Universit\'e de Strasbourg, IPHC, 23 rue du Loess, 67037 Strasbourg, France, CNRS, UMR7178, 67037 Strasbourg, France }
\author{J.~Dudouet}
\affiliation {Grand Acc\'el\'erateur National d'Ions Lourds (GANIL), CEA/DSM-CNRS/IN2P3, Boulevard H. Becquerel, F-14076, Caen, France}
\author{S.~Franchoo}
\affiliation {Institut de Physique Nucl\'eaire d'Orsay, IN2P3-CNRS, F-91406 Orsay Cedex, France}
\author{A.~Gadea}
\affiliation {Instituto De F\'isica Corpuscular, Consejo Superior de Investigaciones Cient\'ificas-University of Valencia, ES-46071 Valencia, Spain}
\author{A.~Gottardo}
\affiliation {LNL(INFN), Laboratori Nazionali di Legnaro, I-35020 Legnaro, Italy}
\affiliation {Dipartimentito di Fisica dell'Universit\`a and INFN, I-35131 Padova, Italy}
\author{T.~H\"uy\"uk}
\affiliation {Instituto De F\'isica Corpuscular, Consejo Superior de Investigaciones Cient\'ificas-University of Valencia, ES-46071 Valencia, Spain}
\author{B.~Jacquot}
\affiliation {Grand Acc\'el\'erateur National d'Ions Lourds (GANIL), CEA/DSM-CNRS/IN2P3, Boulevard H. Becquerel, F-14076, Caen, France}
\author{A.~Kusoglu}
\affiliation {Department of Physics, Istanbul University, 34134 Istanbul, Turkey}
\author{D.~Lebhertz}
\affiliation {Grand Acc\'el\'erateur National d'Ions Lourds (GANIL), CEA/DSM-CNRS/IN2P3, Boulevard H. Becquerel, F-14076, Caen, France}
\author{G.~Lehaut}
\affiliation {Universit\'e de Lyon, Universit\'e Lyon 1, CNRS-IN2P3, Institut de Physique Nucl\'eaire de Lyon, F-69622 Villeurbanne, France}
\author{M.~Martini}
\affiliation {CEA, DAM, DIF, F-91297 Arpajon, France}
\author{D.~R.~Napoli}
\affiliation {LNL(INFN), Laboratori Nazionali di Legnaro, I-35020 Legnaro, Italy}
\author{F.~Nowacki}
\affiliation {Universit\'e de Strasbourg, IPHC, 23 rue du Loess, 67037 Strasbourg, France, CNRS, UMR7178, 67037 Strasbourg, France }
\author{S.~P\'eru}
\affiliation {CEA, DAM, DIF, F-91297 Arpajon, France}
\author{A.~Poves}
\affiliation {Departamento de F\'isica Te\'orica e IFT-UAM/CSIC, Universidad Aut\'onoma de Madrid, E-28049 Madrid, Spain}
\author{F.~Recchia}
\affiliation {Dipartimentito di Fisica dell'Universit\`a and INFN, I-35131 Padova, Italy}
\author{N.~Redon}
\affiliation {Universit\'e de Lyon, Universit\'e Lyon 1, CNRS-IN2P3, Institut de Physique Nucl\'eaire de Lyon, F-69622 Villeurbanne, France}
\author{E.~Sahin}
\affiliation {LNL(INFN), Laboratori Nazionali di Legnaro, I-35020 Legnaro, Italy}
\author{C.~Schmitt}
\affiliation {Grand Acc\'el\'erateur National d'Ions Lourds (GANIL), CEA/DSM-CNRS/IN2P3, Boulevard H. Becquerel, F-14076, Caen, France}
\author{M.~Sferrazza}
\affiliation {D\'epartement de Physique, Universit\'e Libre de Bruxelles, Facult\'e des Sciences, Boulevard du Triomphe, 1050 Bruxelles, Belgium}
\author{K.~Sieja}
\affiliation {Universit\'e de Strasbourg, IPHC, 23 rue du Loess, 67037 Strasbourg, France, CNRS, UMR7178, 67037 Strasbourg, France }
\author{O.~Stezowski}
\affiliation {Universit\'e de Lyon, Universit\'e Lyon 1, CNRS-IN2P3, Institut de Physique Nucl\'eaire de Lyon, F-69622 Villeurbanne, France}
\author{J.J.~Valiente-Dob\'on}
\affiliation {LNL(INFN), Laboratori Nazionali di Legnaro, I-35020 Legnaro, Italy}
\author{A.~Vancraeyenest}
\affiliation {Universit\'e de Lyon, Universit\'e Lyon 1, CNRS-IN2P3, Institut de Physique Nucl\'eaire de Lyon, F-69622 Villeurbanne, France}
\author{Y.~Zheng}
\affiliation {Grand Acc\'el\'erateur National d'Ions Lourds (GANIL), CEA/DSM-CNRS/IN2P3, Boulevard H. Becquerel, F-14076, Caen, France}
\affiliation {Institute of Modern Physics, Chinese Academy of Sciences, 509 Nanchang Rd., Lanzhou, China}

\date{\today}

\begin{abstract}
We report on the observation of a new isomeric state in $^{68}$Ni. We suggest that the newly observed state at 168(1) keV above the first 2$^+$ state is a $\pi(2p-2h)$ 0$^{+}$ state across the major Z=28 shell gap. Comparison with theoretical  calculations indicates a pure proton intruder configuration and the deduced low-lying structure of this key nucleus suggests a possible shape coexistence scenario involving a highly deformed state. 

\end{abstract}


\maketitle


The atomic nucleus is a complex quantum system consisting of two kinds of strongly-interacting fermions.  A direct consequence of this fermionic nature, the Pauli principle, is
the shell  model of the nucleus, one property of which being the existence of magic gaps.  Shell structures are present in a number of systems like atoms, metal clusters, or
quantum dots and wires for instance and are strongly linked to the symmetries of the mean-field. How the shell gaps evolve in nuclei that are further and further away from
stability is one of the key questions to which the  radioactive beam facilities that are currently under construction have to bring answers. Already today, the structure of
moderately exotic nuclei such as $^{68}$Ni allows one to pave the way towards a general answer to the problem of shell evolution. Unusual configurations which are expected to
dominate in the ground state structure of very exotic nuclei can be identified as excited structures in systems not very far away from stability. The strong contribution of the
spin-orbit term in the nucleon-nucleon interaction affects in a major way the single-particle levels with the largest angular momentum, pushing it down in energy. This quenches
significantly the N=40 magic gap from the spherical  harmonic oscillator. The intrusion of the $1g_{9/2}$ and the $2d_{5/2}$ neutron orbitals bring collectivity and enhances
neutron pair excitations across N=40 from the $fp$ shell into the $1g_{9/2}$. Conversely however, this parity change hinders quadrupole excitation and mimics some properties
usually  associated to magicity. In $^{68}$Ni, the observation of a first excited 0$^+_2$ state at low energy \cite{bernas_magic_1982} and the high excitation energy of the
2$^+_1$ state \cite{broda_n40_1995} are examples of such properties. These competing consequences of shell quenching make of $^{68}$Ni a particularly suited case to study the
evolution of shell gaps with isospin. 

Reactions involving single proton particle-hole excitations, $\pi$(1p-1h), are an ideal tool to learn about the residual interaction. Unfortunately they lie at very high
excitation energy. One possibility to circumvent this reef is to look for $\pi$(2p-2h) states which are lowered in energy thanks to pairing correlations and proton-neutron
residual interactions. Studying pair excitation across magic gaps means, therefore, studying these residual interactions. Pair excitations are revealed by the presence of excited
0$^+$ states. In $^{68}$Ni, two such states are reported, mainly of neutron character, originating from the scattering of pairs into the $\nu 1g_{9/2}$. State
corresponding to the excitation of two protons (2p-2h) has been predicted by Pauwels {\it et al.} \cite{pauwels_shape_2008,pauwels_pairing-excitation_2010} using the energy of
the intruder $\pi$(2p-1h) state in $^{69}$Cu and, symmetrically, the $\pi$(1p-2h) in $^{67}$Co, which both lie at N=40. The energy they derive leads to a low value of 2202 keV,
which can be understood only with an important gain  in binding energy from the $\pi-\nu$ residual interactions between the two proton-holes and the active valence  neutrons
across N=40 \cite{pauwels_shape_2008,pauwels_pairing-excitation_2010}. The spin-parity of the $\pi$(1p-2h) state in $^{67}$Co was proposed to be (1/2$^{-}$) and corresponding to a prolate proton intruder configuration~\cite{pauwels_shape_2008}. This shape isomer indicates therefore the presence of deformed low-lying proton intruder states below $^{68}$Ni. Later on, evidences for such deformed ($\pi$p$_{3/2}$) intruder orbital were reported in the spectroscopy of odd mass Mn isotopes~\cite{Liddick2011_1} in $\beta$-decay experiments. 
These results strongly suggest the presence of a deformed $\pi$(2p-2h) intruder state in $^{68}$Ni. But despite the large number of experiments dedicated to $^{68}$Ni, no evidence of proton-pair excitation has been reported so far, preventing us from a coherent understanding of the nuclear structure in this mass region. In the present work, we report on the observation of a new isomeric state in $^{68}$Ni which we propose to have a 0$^{+}$ character.

The experiment was performed at the Grand Acc\'el\'erateur National  d'Ions Lourds (GANIL) by using multi-nucleon transfer reactions in inverse kinematics. A $^{238}$U beam at
6.33 A.MeV bombarded a 1.3 mg/cm$^{2}$ thick $^{70}$Zn target.  The target-like reaction products were detected and identified in the VAMOS
spectrometer~\cite{savajols_vamos:_2003,pullanhiotan_performance_2008} used  in a soleno\"id mode. The optical axis of the spectrometer was set at $45^\circ$ with respect to the
beam axis, such that the grazing angle was within the angular  acceptance of the spectrometer. In this mode, the  dipole of the spectrometer is not used and the transmission is increased by
50\% with respect to the standard dispersive mode. The reaction products are refocused in a new detection setup \cite{dijon_NIMA} located at the focal plane of the spectrometer, 
which provides an unambiguous identification of the recoils on an event-by-event basis.  The atomic number is measured by combining the energy loss in three successive ionization
chambers  and the residual energy deposited in four silicon detectors. The mass is determined from the total kinetic energy and the time of flight between silicon detectors and a
Multi Wire  Proportional Chamber located 138 mm downstream the target. The flight path through the spectrometer is determined using Secondary Electron Detectors
\cite{Drouart20071090}  and following the procedure described in \cite{savajols_vamos:_2003}. We have obtained a mass resolution  $\Delta A/A$ = 1.2\% and an atomic number
resolution $\Delta Z/Z$=1.1\%. Prompt $\gamma$-rays emitted at  the target position were measured by eleven clover detectors from the EXOGAM array~\cite{simpson_exogam_2000} in
coincidence with the recoils identified in VAMOS. Delayed $\gamma$-rays  were also detected at the VAMOS focal plane by four HPGe detectors facing the silicon detectors where the 
recoils were implanted. With a recoil velocity from 22 to 45 $\mu$m/ps, the flight time through the spectrometer  was 200-400 ns. The lifetimes have been measured using a time to
digital conversion module with a hardware  gate of 3 $\mu$s. Our delayed spectroscopy setup is therefore sensitive to lifetimes from $\sim$100 ns up to $\sim$10 $\mu$s. 

\begin{figure}
\centering
\includegraphics[width=8.7cm,height=4.7cm]{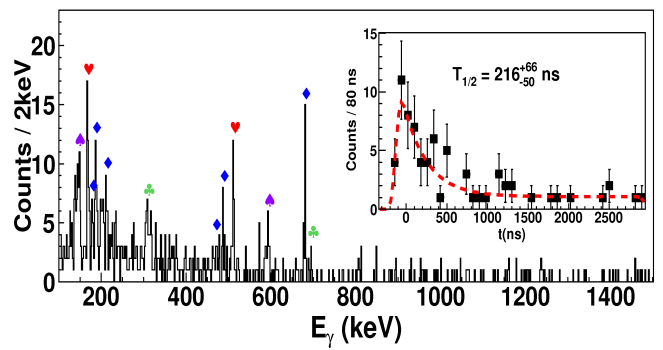}
\caption{(Color online) Delayed $\gamma$-ray spectrum in coincidence with $^{68}$Ni. The purple spades, the blue diamonds and the green 
clovers indicate respectively the $^{69}$Ni, $^{69}$Cu and $^{67}$Ni isomeric decay. The red heart 
overscores the transition that belongs to $^{68}$Ni. The inset shows the time spectrum from which 
the half-life of the 168(1) keV transition is extracted.}
\label{GammaResult}
\end{figure}

In this Rapid Communication, we focus on new results obtained on the $^{68}$Ni key nucleus. Fig. \ref{GammaResult} shows the delayed $\gamma$-ray spectrum observed in coincidence with its
identification in VAMOS. Several isomers are known in $^{68}$Ni: an 8$^+$ seniority isomer with a half-life t$_{1/2}$ = 23.3(11) ns  \cite{Ishii_2000}, which decays in-flight
in the spectrometer; a 5$^-$ state with t$_{1/2}$ = 0.86(5) ms \cite{Burrows20021},  which is too long-lived for its decay to be observed; and a $0^+_2$ state with t$_{1/2}$ = 270(5) ns
\cite{Sorlin2002}. Given the mass and charge resolution we measured, the spectrum  obtained after selection of $^{68}$Ni is slightly contaminated by neighboring nuclei,
$^{67,69}$Ni and $^{69}$Cu, in which isomers are known. All the transitions from their decay have been identified (see Fig.\ref{GammaResult}). In particular, the transitions
observed in $^{69}$Cu correspond to the decay of a $13/2^{+m}$ state. Its half-life is well known (t$_{1/2}$ = 360(30) ns) \cite{Ishii_2000} and allowed us to fully control  our setup and
procedures. Our measurement yields the value of t$_{1/2}$ = 360(20) ns in excellent agreement.

\begin{figure}
\centering
\includegraphics[width=8cm,height=8cm]{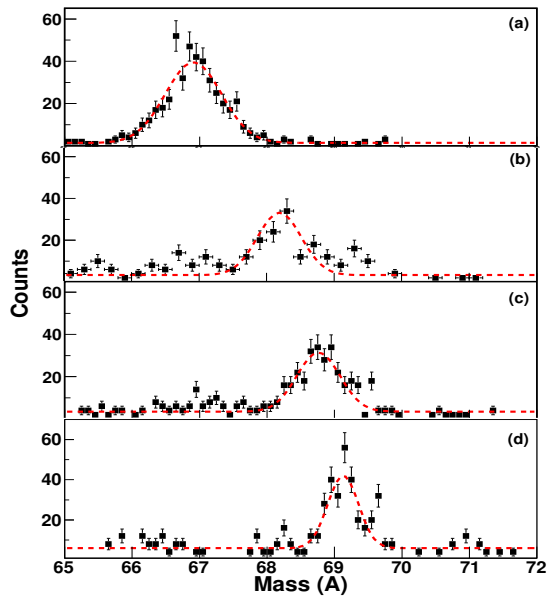}
\caption{(Color online) Mass spectra in coincidence with the observed delayed $\gamma$-rays. 
The coincidence gates select: (a) the 313+694 keV transitions in $^{67}$Ni, (b) the 168 keV transition, 
(c) the $^{69}$Cu lines and (d) the 143+593 keV transitions in $^{69}$Ni.}
\label{MassResult}
\end{figure}

In addition, two peaks at 168 keV and 511 keV are clearly visible. The procedure to assign these lines to a given nucleus is described in Fig. \ref{MassResult}. This figure shows
the mass spectrum measured by VAMOS and in coincidence with the observed delayed $\gamma$-rays. The mass distribution given by the coincidence with the 168 keV line is A =
68.19(4) with $\sigma$ = 0.32(4) i.e. 68. This procedure applied to the Z distribution confirms the assignment to nickel. The same is true for the 511 keV line.  Therefore, they
both have been assigned to $^{68}$Ni. The 511 keV transition most probably arises from the internal pair creation in the decay of  the 0$^+_2 $ to the ground state. No isomeric
state decaying by a 168(1) keV line was reported so far in $^{68}$Ni. The half-life of the new state decaying via the 168 keV transition was measured at t$_{1/2}$ =
216($^{+66}_{-50}$) ns as shown in the inset of Fig. \ref{GammaResult}.  No other transition is observed.

It is unfortunately not possible to use $\gamma$-$\gamma$ coincidences at the focal plane to build the level scheme. The non-observation of the 2$^+_1 \rightarrow 0^+_1$
transition is consistent with a decay of the isomeric state to the 2$^+_1$ state due to the very low cross-section to populate this new isomer (as given by the statistics in the
168 keV peak) and to the low efficiency of the four Ge detectors at the focal plane at 2 MeV: the efficiency ratio between  168 keV and 2 MeV is 16, which leads to less than 3
counts in the 2 MeV peak assuming a 100\% decay to the  $2^+_1$ state. This is below our detection limit. All the other known transitions in $^{68}$Ni have an energy lower than 2
MeV and should have been observed if lying in coincidence with the isomer decay. Weisskopf estimates for the half-life of an E1, E2, M1 and M2  168 keV transition are 0.1 ps, 300
ns, 6 ps and 20 $\mu$s respectively.  The measured half-life indicates therefore an E2 character. The corresponding very low B(E2$\downarrow$) $\leq$ 23.5 $e^2fm^4$ ( $\leq$ 1.4 
W.u.) (assuming a pure E2 transition), indicates a single particle character. Spin and parity of the new isomeric state can be inferred from the direct transfer reactions
$^{70}$Zn($^{14}$C,$^{16}$O) performed in \cite{bernas_magic_1982,girod_spectroscopy_1988}. Two states assigned to $^{68}$Ni  have been measured at 1770(30) keV and 2200(30) keV.
The angular distributions and  Distorted Wave Born Approximation calculations for the first excited state  gave an unambiguous 0$^+_2$ assignment. The situation for the second
excited state  was much more uncertain due to the presence of a contaminant peak at E $\sim 2.3$ MeV  arising from the presence of oxygen in the target. For this reason, the two
data points closest to  0$^{\circ}$ scattering angle were removed from the analysis and the authors tentatively  identified this state as the 2$^+_1$, not observed at that time,
but later measured  at 2033 keV \cite{broda_n40_1995}. The 2.2 MeV state might well be the one we report in this letter. In fact, it is one of the properties of the angular
distributions of the 0$^+$$\rightarrow$0$^+$ to be strongly forward peaked. It is also clear that in \cite{girod_spectroscopy_1988}, the angular distribution could also be fitted
as a $\Delta$L=0 transfer. Consequently, we tentatively assign this E$_\gamma$=168(1) keV transition to the decay of a new (0$^{+m}_3$) state at 2202(1) keV to the 2$^+_1$. The position of the
(0$^{+m}_3$) level is in remarkable agreement with the 2202 keV $\pi$(2p-2h) 0$^+$ intruder state, i.e. ($1\pi f_{7/2}$)$^{-2}$, deduced by Pauwels {\it et al.}  \cite{pauwels_pairing-excitation_2010}. The
excitation energy at the nearly exact sum of the $\pi$(2p-1h) and $\pi$(1p-2h) intruder state excitation energy in $^{69}$Cu and $^{67}$Co respectively is consistent with the fact that the wave function has a pure $\pi$(2p-2h) character. This further supports the single particle character deduced from our measured E2 transition rate. Any
possible E0 decay would strengthen even more this conclusion.

The $^{68}$Ni nucleus has been produced and studied by various means: transfer reactions, as already discussed, but also using deep-inelastic reactions, fragmentation
reactions and $\beta$-decay studies. The question of the non observation of the 168 keV transition in these earlier works
then arises. We carefully examined the previous studies known to us, and it first turns out that none of the deep-inelastic experiments
\cite{broda_n40_1995,Ishii_2000,Ishii_2002} was suited to possibly detect it (no adequate isomer setup, high gamma-ray-fold hardware trigger or high-lying transition
gate in the analysis). Fragmentation reactions dedicated to the quest of new isomers have been performed in particular at the LISE fragment separator. 
In these experiments, and when the flight path of the fragments was compatible with the new isomer half-life, neither the decay from the $8^{+}$ seniority
isomer nor the one from the new ($0_{3}^{+m}$) state is observed. This has been checked in data from the fragmentation of an $^{86}$Kr beam~\cite{Daugas_phd}. It is clear from these data that the decay of $^{68}$Ni proceeds essentially via the $5^{-}_{1}$ and the $0^{+}_{2}$ isomers. This supports the conclusion of~\cite{Daugas_2} showing that the feeding pattern of isomers from
intermediate energy fragmentation is a complex interplay between reaction channels, beam energy and
momentum distribution selected by the spectrometer. 

Finally, $^{68}$Ni was populated by $\beta$-decay of $^{68}$Co~\cite{mueller__2000}, which has a (7$^-$) ground state. From the $\beta$-decay selection rules, it is clear that the 7$^-$$\rightarrow$ 0$^+$ transition is highly forbidden. Another isomer in $^{68}$Co has also been observed to decay in $^{68}$Ni which has been attributed a (3$^{+}$) character in~\cite{Bosch198889,mueller__2000} with a ($\pi$f$_{7/2}$)$^{-1}$($\nu$p$_{1/2}$)$^{-1}$($\nu$g$_{9/2}$)$^{+2}$ configuration. A small fraction of the flux is feeding a state which has been tentatively assigned a (0$^+$) character. However, this decay would be a doubly-forbidden Gamow-Teller transition. In addition, the $\pi f^{-1}_{7/2} \rightarrow \nu f^{-1}_{5/2}$ transition from the (3$^{+}$) state populates preferentially neutron states in $^{68}$Ni as observed in \cite{mueller__2000}. More recently, in~\cite{Liddick2011_2}, it has been proposed to reassign from (3$^{+}$) to (1$^{+}$) the spin and parity of this isomer with however some inconsistency between apparent $\beta$-decay feeding of the first excited state at 45 keV and its E1 decay to the isomer. The non observation of the 168 keV line in $\beta$-decay of $^{68}$Co$^{m}$ would rather confirm the neutron character of the isomer hence the (3$^{+}$) spin and parity. It is therefore very unlikely that the newly discovered state is fed by $\beta$-decay. Another possibility that cannot be excluded however is a large fragmentation of the wave function of the (0$^{+m}_{3}$) in $^{68}$Ni yielding to a very weak population, i.e. below the sensitivity limit~\cite{Liddick2011_2}. 
\begin{figure}
\centering
\includegraphics[width=8.5cm,height=7.5cm]{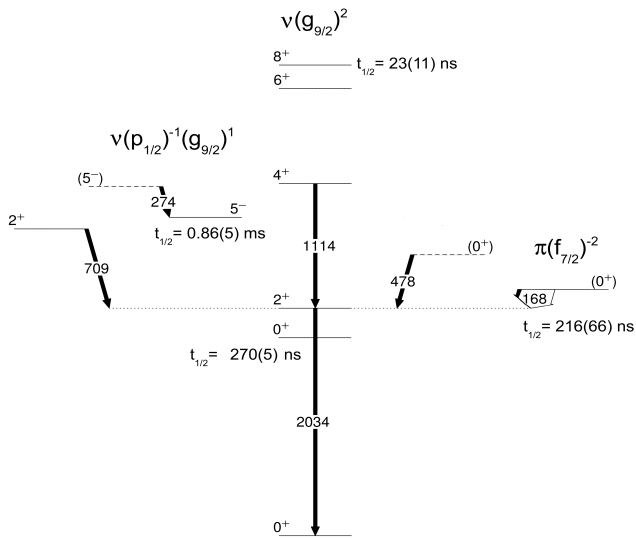}
\caption{Proposed level scheme deduced from the present work in $^{68}$Ni. The open (filled) arrows indicate respectively the observed delayed (prompt) $\gamma$-ray. The
configurations which are discussed in the text and isomers half-life are indicated.}
\label{LevelResult}
\end{figure}

The shell evolution of N=40 nuclei presents many similarities with the N=20 region around $^{32}$Mg and was presented as a new  {\it island of inversion} \cite{ljungvall_onset_2010}. The large scale shell
model (LSSM) calculations performed with the most recent interaction (LNPS) tailored for the mass region and in an extended valence space \cite{lenzi_2010} show that the strong deformation in the Cr chain is
the result of the proton-neutron correlations between the neutrons populating the  $1g_{9/2}$-$2d_{5/2}$ orbitals and the valence protons in the $pf$ shell. This interaction satisfactory reproduces experimental data in the mass region \cite{lenzi_2010}. We performed LSSM calculations using the LNPS interaction for $^{68}$Ni. In the calculations, a third 0$^+_3$ state \cite{lenzi_2010} is found
at 2.4 MeV (i.e. very close to the proposed (0$^{+m}_3$)) for which the dominant configuration is a $\pi$(2p-2h) i.e. ($1\pi f_{7/2}$)$^{-2}$. This {\it intruder} state has only {\it normal} parity neutron states as neighbors. The 2$^+_1$ wave function is
calculated to be a $\nu$(2p-2h) configuration and hence with a negligible overlap with the theoretical 0$^{+}_3$ state. As a result, the calculated B(E2$\downarrow$) is 0.9 W.u in good agreement with the experimental
value. The correlation energies, quadrupole sum rule and fragmentation of the calculated wave functions reflect the difference in nature of low lying $0^+$ states. The spherical ground state is characterized 
by -7.6MeV of correlation energy, 899~$e^2fm^4$ quadrupole sum rule and 61\% of closed shell configuration, whereas for the proton intruder $0_3^+$ we obtain -30.9MeV of correlation energy, 4520~$e^2fm^4$
\footnote{Carrying the calculations for the present letter, it was found that the sum rule indicated in \cite{lenzi_2010} is {\it not} correct and should be replaced by the present value of 4520~$e^2fm^4$ for
this $0_3^+$ state } of quadrupole sum rule and a wave function spread out over many spherical components. 
The calculated intrinsic quadrupole moment $Q_{int}\sim~195~efm^{2}$ corresponds to a very large quadrupole deformation of $\beta_2\sim$ 0.4, a value which is comparable to the one measured in superdeformed bands of the zinc region~\cite{Svensson1997}. In addition, the second $2^{+}$ level is calculated just 234 keV above the $0^{+}_{3}$ state. They have very similar proton and neutron configurations and the $B(E2;2^{+}_{2} \rightarrow 0^{+}_{3})$ is as large as 46 W.u., which confirms the highly collective character of the band built on top of the $0^{+}_{3}$ state. 

Complementary, the collective structures in N=40 isotones have been recently investigated within the Hartree-Fock-Bogoliubov approach using the Gogny D1S effective interaction \cite{gaudefroy_2009}. Coupling between collective and non-collective degrees of freedom are not included in the approach and therefore  any two-proton state is beyond reach of the model. However, one can extract hints on the  deformation from the calculated pairing energies. In $^{68}$Ni, proton and neutron pairing energies vanish simultaneously  at large $\beta$ deformation ($\beta\sim$ 0.45) strongly suggesting a possible local minimum in the potential energy surface, hence a shape coexistence scenario involving a highly deformed state. The (0$^{+m}_3$) state is obviously a natural candidate for such a minimum. 

Finally, fully consistent QRPA calculations with D1S interaction \cite{QRPA2008} were performed for $^{68}$Ni. Three excited 0$^+$ states at 2.14 MeV, 2.24 MeV and 3.47 MeV respectively are predicted. The two first states most probably correspond to a single physical state at 2.14 MeV. This is due to the approximation used in the calculation of 2p-2h states in QRPA. This state is absent in the corresponding ph-RPA calculations and it has a pure proton character. Its wave function is dominated by the $(1\pi f_{7/2})^{-2}$ configuration in excellent agreement with the proposed interpretation. The 0$^+$ excited state at 3.47 MeV has a pure neutron configuration and might correspond to the first experimental excited state.

Both experimental and theoretical investigations give, therefore, a more coherent picture of the low-lying structure of $^{68}$Ni where the 0$^+_{2,4}$ result from the scattering of pairs into the $1\nu g_{9/2}$ and the (0$^{+m}_{3}$) is interpreted as a ($1\pi f_{7/2}$)$^{-2}$ configuration (see Fig. \ref{LevelResult}). Many examples of such $\pi(2p-2h)$ 0$^{+}$ states are known in semi-magic nuclei \cite{Wood1992101}. They occur at lowest excitation energy when the number of active valence nucleons is at mid shell (e.g., around $^{116}$Sn or $^{186}$Pb) because neutron-proton correlations are maximal at that point. What makes $^{68}$Ni  unique is that, despite the doubly-magic character of the dominant component of its ground state, it has a low-lying $\pi(2p-2h)$ 0$^{+}$ state as a result of an unusually large neutron-proton correlation energy.
The configuration of the newly observed (0$^{+m}_3$) state with active protons in the $pf$ shell is very similar to the Fe case, i.e. with two holes in the $1\pi f_{7/2}$ orbital. Therefore one can reasonably consider that the configuration corresponding to the (0$^{+m}_3$) state at 2202(1) keV in $^{68}$Ni migrates down to 491 keV excitation energy in $^{67}$Co to become the ground state in $^{66}$Fe. The {\it normal} configuration, as opposed to the {\it intruder}  one, might then appear as a low-lying 0$^+$ state in $^{66}$Fe. 

In summary, a new isomeric state has been observed at 168(1) keV above the first 2$^+$ state in $^{68}$Ni. It is interpreted as the intruder $\pi$(2p-2h) 0$^{+}$  state across the major Z=28 shell gap. This interpretation is supported by large scale shell model calculations which also indicate a highly deformed state with $\beta_2\sim$ 0.4. Our observation fits extremely well with the prescription used in \cite{pauwels_pairing-excitation_2010} to predict the energy of the $\pi$(2p-2h) and also with the most recent theoretical calculations, indicating a pure proton excitation character for this state. The systematics of intruder states along the N=40 isotonic chain with Z$\leq$28 strongly suggests that shape coexistence is occurring at low energy in $^{68}$Ni.

We would like to thank P. Van Isacker for stimulating discussions. A. G. and G. de F. acknowledge the support of IN2P3, France, and MICINN,  Spain, through the AIC10-D-000429 bilateral action.  A. G. and T.
H.  activity has been partially supported by the MICINN and Generalitat  Valenciana, Spain, under grants FPA2008-06419 and
PROMETEO/2010/101. Z. Y. acknowledges support of the Chinese Academy of Science, China.

\bibliographystyle{apsrev4-1}
\bibliography{Nickel68_PRL_Dijon}

\end{document}